\documentclass[twocolumn]{aastex631}

\usepackage{amsmath}
\usepackage[flushleft]{threeparttable}

\shorttitle{Planet mass function at 1-10 au}
\shortauthors{Chachan \& Lee} 

\begin{document}

\title{Planet Mass Function around M stars at 1-10 au: A Plethora of sub-Earth mass objects}

\correspondingauthor{Yayaati Chachan}
\email{yayaatichachan@gmail.com}

\author[0000-0003-1728-8269]{Yayaati Chachan}
\altaffiliation{CITA National Fellow}
\affiliation{Department of Physics and Trottier Space Institute, McGill University, 3600 rue University, H3A 2T8 Montreal QC, Canada}
\affiliation{Trottier Institute for Research on Exoplanets (iREx), Universit\'e de Montr\'eal, Canada}
\affiliation{Department of Astronomy and Astrophysics, University of California, Santa Cruz, CA 95064, USA}

\author[0000-0002-1228-9820]{Eve J.~Lee}
\affiliation{Department of Physics and Trottier Space Institute, McGill University, 3600 rue University, H3A 2T8 Montreal QC, Canada}
\affiliation{Trottier Institute for Research on Exoplanets (iREx), Universit\'e de Montr\'eal, Canada}

\begin{abstract}
Small planets ($\lesssim 1$ M$_\oplus$) at intermediate orbital distances ($\sim$1 au) represent an uncharted territory in exoplanetary science. The upcoming microlensing survey by the \emph{Nancy Grace Roman Space Telescope} will be sensitive to objects as light as Ganymede and unveil the small planet population at $1-10$ au. Instrumental sensitivity to such planets is low and the number of objects we will discover is strongly dependent on the underlying planet mass function. In this work, we provide a physically motivated planet mass function by combining the efficiency of planet formation by pebble accretion with the observed disk mass function. Because the disk mass function for M dwarfs (0.4--0.6$M_\odot$) is bottom heavy, the initial planet mass function is also expected to be bottom-heavy, skewing towards Ganymede and Mars mass objects, more so for heavier initial planetary seeds. We follow the subsequent dynamical evolution of planetary systems over $\sim$100 Myr varying the initial eccentricity and orbital spacing. For initial planet separations of $\geq$3 local disk scale heights, we find that Ganymede and Mars mass planets do not grow significantly by mergers. However, Earth-like planets undergo vigorous merging and turn into super-Earths, potentially creating a gap in the planet mass function at $\sim 1$ M$_{\oplus}$. Our results demonstrate that the slope of the mass function and the location of the potential gap in the mass function can probe the initial architecture of multi-planet systems. We close by discussing implications on the expected difference between bound and free-floating planet mass functions.
\end{abstract}

\section{Introduction} \label{sec:intro}

The abundance of small planets ($\lesssim 1$ M$_\oplus$) remains a terra incognita in the field of exoplanetary science. Given their low masses and radii, they are incredibly difficult to find using techniques that rely on a planet's influence on its host star. Consequently, planets akin to the terrestrial planets in the solar system have only been found around other stars on extremely short orbits. Their distribution is largely unknown even at Mercury-like distances (0.3 au, $\sim 100$ days) and it is completely unknown for distances $\gtrsim 1$ au \citep[e.g.,][]{Hsu19, Hsu20, Dattilo2023, Zhu2021}. Answering the question of whether our solar system is common or not therefore remains largely out of reach. 

Fortunately, a dramatic change in our knowledge of small planets at large orbital distances is imminent. The upcoming \emph{Nancy Grace Roman Space Telescope} will open new frontiers in the characterization of exoplanet populations at distances of $1-10$ au. Using the microlensing technique, which detects planets that gravitationally lens background stars and cause a brief jump in their brightness \citep{Paczynski1986, Gould1992}, it will enable us to detect objects as small as Ganymede \citep{Penny2019, Zhu2021}! Hitherto, the lowest mass planets that microlensing surveys from the ground have managed to detect are super-Earths \citep[e.g.,][]{Zang2023}. The impact of pushing the detection limit to planets that are two orders of magnitude less massive will likely be similar to that of the \emph{Kepler Space Telescope} more than a decade ago. 

The number of small planets that we expect to detect with the \emph{Nancy Grace Roman Space Telescope} is heavily dependent on how abundant such planets are, i.e. the planet mass function. Previous planet yield estimates have used constant or power law planet occurrence rate models for low mass planets \citep{Penny2019} that either extrapolate from ground-based microlensing detections \citep{Cassan2012} or inferred from {\it Kepler} planets assuming a fixed mass-radius relation \citep{Lissauer2011}.
The aim of our study is to provide a physically motivated planet mass function at $1-10$ au for planets less massive than super-Earths. In particular, we quantify the planet mass function expected at birth as a result of pebble accretion (\S~\ref{sec:formation}). Using pebble accretion efficiency and the measured disk mass function, we calculate the fraction of stars that host low mass planets in \S~\ref{sec:planet_mass_dist}. 

Dynamical evolution over long timescales can alter the natal planet mass function of both bound and unbound (`free-floating') planets. We perform a suite of N-body simulations for low mass planets to characterize the effects of dynamical evolution on the planet mass function for a range of initial conditions. The details and results of the simulations are presented in \S~\ref{sec:planets_evolve}. Finally, in \S~\ref{sec:discussion}, we summarize our results and discuss the implications of our work for observations by the upcoming \emph{Nancy Grace Roman Space Telescope}.

\section{Formation scenario and model}
\label{sec:formation}
We hypothesize that planetary seeds emerge in protoplanetary disks as a result of some solid clumping (e.g., streaming instability, \citealt{Johansen17, Li2021}) by the beginning of the Class I stage. These seeds then grow by accreting the remaining mm-cm sized dust (pebbles) that is drifting past them. The final mass to which these seeds grow depends on the amount of dust present in the disk and the efficiency with which this dust is accreted. 

The efficiency $\epsilon$ with which the drifting pebbles are accreted is given by the ratio of the pebble accretion rate and the rate at which pebbles drift
\begin{equation}
    \epsilon = \frac{\dot{M}_{\rm peb}}{\dot{M}_{\rm drift}}.
\end{equation}
Detailed calculation of $\epsilon$ is outlined in \citet{Chachan23}, which we adopt. Here, we only highlight the key results.
The drift rate of pebbles $\dot{M}_{\rm drift} = 2 \pi v_{\rm r} \Sigma_{\rm d, St}$, where $v_{\rm r}$ is the radial velocity of the pebbles and $\Sigma_{\rm d, St}$ is the surface density of pebbles with Stokes number St. The radial velocity of pebbles is the sum of coupled radial motion due to the gas and the radial drift of pebbles due to slightly sub-Keplerian velocity of the pressure-supported gas:
\begin{equation}
    v_{\rm r} =  -\frac{3}{2}\frac{\nu}{r}\frac{1}{1+{\rm St}^2} - 2\eta v_{\rm K} \frac{\rm St}{1+{\rm St}^2},
\end{equation}
where $r$ is the radial distance from the star, $\nu = \alpha_{\rm t} c_{\rm s} H_{\rm g}$ is the kinematic viscosity of the gas, $\alpha_{\rm t}$ is the Shakura-Sunyaev parameter, $c_{\rm s}$ and $H_{\rm g} = c_{\rm s} / \Omega_{\rm K}$ are respectively the local sound speed and scale height of the gas, $\Omega_{\rm K}$ is the Keplerian frequency, $v_{\rm K} = \Omega_{\rm K} r$ is the Keplerian velocity, and the negative sign indicates inward motion towards the central star. The quantity $\eta = -0.5 \gamma (c_{\rm s} / v_{\rm K})^2$ quantifies the deviation of gas' motion from Keplerian and $\gamma = d{\rm ln}P_{\rm g}/d{\rm ln}r$ is the logarithmic pressure gradient. 

The pebble accretion rate $\dot{M}_{\rm peb}$ is given by
\begin{equation}
    \dot{M}_{\rm peb} = 2\Sigma_{\rm d, St} R_{\rm acc} v_{\rm acc} \times {\rm min}(1, R_{\rm acc}/H_{\rm d}),
\end{equation}
where pebbles traveling at relative speed of $v_{\rm acc}$ are accreted onto the planet if they approach closer than $R_{\rm acc}$, and $H_{\rm d} \equiv H_{\rm g} [\alpha_{\rm t} / (\alpha_{\rm t} + {\rm St})]^{1/2}$ is the scale height of the pebble disk. We then have
\begin{align}
&\dot{M}_{\rm peb}\sim \nonumber \\
&\begin{cases} 
    (192)^{1/3}{\rm St}^{2/3}q^{2/3}\Omega_{\rm K}\Sigma_{\rm d, St}r^2 & \hspace{.1cm} (\text{2D, hw})\\
    4\eta^{1/2}{\rm St}^{1/2}q^{1/2}\Omega_{\rm K}\Sigma_{\rm d, St}r^2 & \hspace{.1cm} (\text{2D, sh})\\
    \dfrac{8\Sigma_{\rm d, St}GM_{\rm p}{\rm St}}{c_s}\left(\dfrac{\alpha_{\rm t} + {\rm St}}{\alpha_{\rm t}}\right)^{1/2} & \hspace{.1cm} (\text{3D})
\end{cases}
\label{eq:peb_acc_rate}
\end{align} 
where $G$ is the gravitational constant, $M_{\rm p}$ is the protoplanet mass, and $q = M_{\rm p} / M_{\star}$ is the protoplanet mass ratio. Here, (2D, hw) case refers to $R_{\rm acc} > H_d, v_{\rm acc} = \eta v_{\rm k}$ (headwind); (2D, sh) to $R_{\rm acc} > H_d, v_{\rm acc} = 3\Omega_{\rm k}R_{\rm acc}/2$ (shear); and (3D) to $R_{\rm acc} < H_d$.

Putting the expressions for $\dot{M}_{\rm peb}$ and $\dot{M}_{\rm drift}$ together and for pebbles with St$\lesssim 1$ such that $(1 + {\rm St}^2) \sim 1$, we obtain the following expressions for accretion efficiency in different regimes:
\begin{align}
&\epsilon= \nonumber \\
&\begin{cases} 
    \dfrac{2}{3\pi}\bigg(\dfrac{q{\rm St}}{\eta}\bigg)^{1/2} |\gamma|\bigg(\alpha_t+\dfrac{2}{3}|\gamma|{\rm St}\bigg)^{-1} & \hspace{.1cm} (\text{2D, hw})\\
    \dfrac{(192)^{1/3}}{6\pi}\bigg(q{\rm St}\bigg)^{2/3}\bigg(\dfrac{|\gamma|}{\eta}\bigg)\bigg(\alpha_t+\dfrac{2}{3}|\gamma|{\rm St}\bigg)^{-1} & \hspace{.1cm} (\text{2D, sh})\\
    \dfrac{4}{3\pi}q\dfrac{r}{H_{\rm d}}\dfrac{|\gamma|{\rm St}}{\eta}\bigg(\alpha_t+\dfrac{2}{3}|\gamma|{\rm St}\bigg)^{-1}. & \hspace{.1cm} (\text{3D})
\end{cases}
\label{eq:acc-eff}
\end{align} 
This expression is only valid for planets on circular and coplanar orbits - we neglect the effects of non-zero eccentricity or inclination on $\epsilon$ \citep[e.g., see][]{Ormel18}. Given that there is a range of pebble sizes present in protoplanetary disks, we calculate the mass-averaged accretion efficiency $\bar{\epsilon}$ assuming dust grains are in the Epstein drag regime (St $\propto a \implies m \propto {\rm St}^3$, where $a$ and $m$ are the grain size and mass respectively):
\begin{equation}
    \bar{\epsilon} (M_{\rm p}) = \dfrac{\int \epsilon (M_{\rm p}, {\rm St}) \, n({\rm St}) \, m({\rm St}) \, {\rm d}{\rm St}}{\int  n({\rm St}) \, m({\rm St}) \, {\rm d}{\rm St}},
\end{equation}
with the grain size distribution given by $n({\rm St}) = dn/d{\rm St}$. For most of our parameter space, fragmentation limits the maximum grain size to Stokes number St$_{\rm frag} = v_{\rm f}^2 / 3 \alpha_{\rm t} c_{\rm s}^2$, where $v_{\rm f}$ is the fragmentation velocity of dust grains \citep{Brauer2008}. The size distribution $n({\rm St})m({\rm St}) \propto {\rm St}^{-0.5}$ for St $<2\alpha_{\rm t}/\pi$ (turbulent regime) and $n({\rm St})m({\rm St}) \propto {\rm St}^{-0.75}$ for St $>2\alpha_{\rm t}/\pi$ (settling regime) \citep{Birnstiel11}. Radial drift is the limiting process only at large distances ($\gtrsim 5$ au) when St$_{\rm frag}$ is large ($\alpha_{\rm t} = 10^{-4}$, $v_{\rm f} = 10$ m s$^{-1}$). For radial drift dominated regime, the maximum Stokes number St$_{\rm drift} = \delta v_{\rm K}^2 / |\gamma| c_{\rm s}^2$, where $\delta = 0.01$ is the local dust-to-gas mass ratio. The size distribution is more top-heavy than in the fragmentation dominated regime with $n({\rm St})m({\rm St}) \propto {\rm St}^{0.5}$ (this corresponds to $n(a) \propto a^{-2.5}$, \citealt{Birnstiel2024}). We integrate $\epsilon$ over St in the range [$10^{-6}$,(min(St$_{\rm frag}$, St$_{\rm drift}$)] to obtain $\bar{\epsilon}$, where the lower limit is chosen so as to capture grains that are well coupled to the gas and that move radially due to gas' radial motion.

\begin{figure*}
    \centering
    \includegraphics[width=0.50\linewidth]{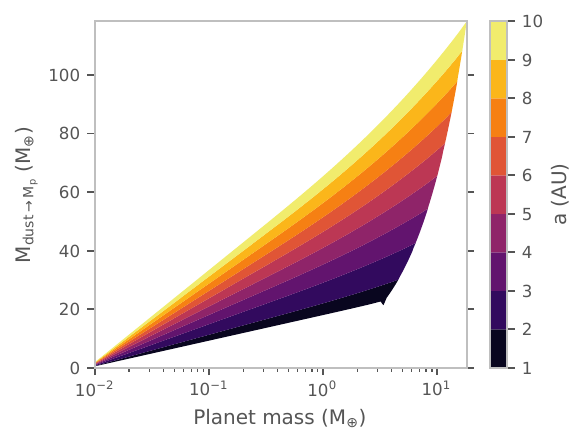}
    \includegraphics[width=0.49\linewidth]{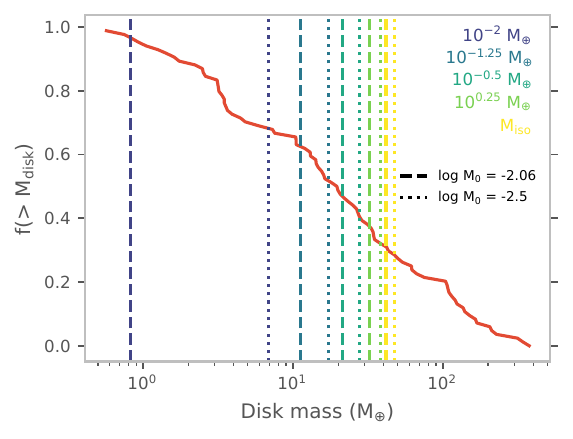}
    \caption{Left panel: The amount of dust mass needed to grow planets ($M_{\rm dust\rightarrow M_{\rm p}}$) from $10^{-2.06} M_\oplus$ as a function of final planet mass $M_{\rm p}$ at a given orbital distance indicated by color.
    Right panel: The cumulative distribution function (CDF) of the observed disk dust mass $M_{\rm disk}$ around stars of 0.4--0.6 $M_\odot$ from \citet{Manara2023} scaled up by a factor of 3 to represent Class 0/I disks \citep{Chachan23, Tychoniec2020}. The vertical lines indicate the dust mass needed to grow a planet seed to a given planet mass at 3 au for two different $M_0$ values, where we choose 3 au to represent the orbital distance to which microlensing surveys are most sensitive \citep{Gaudi2012}. The fraction of stars that can harbor planets in a given mass range is obtained by subtracting the disk mass CDF values corresponding to the planet mass bin edges. Both panels are for fiducial values of $\alpha_{\rm t} = 10^{-4}$ and $v_{\rm f} = 1$ m/s.}
    \label{fig:Mpl_Mdust_relation}
\end{figure*}

Equipped with mass-averaged accretion efficiency $\bar{\epsilon}$, we can calculate the amount of dust mass needed to grow a planet from an initial mass $M_0$ to a given mass $M_{\rm p}$:
\begin{equation}
    M_{\rm dust\rightarrow M_{\rm p}} = \int_{M_0}^{M_{\rm p}} \frac{1}{\bar{\epsilon}} \, dM_{\rm p}.
    \label{eq:mdust_mp}
\end{equation}
This differs from our calculations in \cite{Chachan23} where we calculated the dust mass needed to grow a seed to pebble isolation mass $M_{\rm iso}$ \citep{Bitsch18}:
\begin{multline}
    M_{\rm iso} = 25 M_{\oplus} \bigg(\frac{M_{\star}}{M_{\odot}}\bigg) \bigg(\frac{H_{\rm g}/r}{0.05}\bigg)^3  \bigg[1 - \frac{\gamma + 2.5}{6} \bigg] \\ \times \bigg[0.34 \bigg(\frac{-3}{{\rm log_{10}} \alpha_{\rm t}}\bigg)^4 + 0.66 \bigg].
    \label{eq:pebble_iso}
\end{multline}
Here, we focus on planet mass $M_{\rm p} \leq M_{\rm iso}$. Of particular note and importance in this work is the efficiency's dependence on planet mass. For low mass planets that are the focus of this work, pebble accretion primarily happens in the 3D regime. In the 3D accretion regime, $\epsilon \propto M_{\rm p}$ and as a result, $M_{\rm dust\rightarrow M_{\rm p}} \propto {\rm log}(M_{\rm p} / M_0)$ (see left panel of Figure \ref{fig:Mpl_Mdust_relation}), leading to the difference in $M_{\rm dust\rightarrow M_{\rm p}}$ at regular logarithmic interval of $M_{\rm p}$ being the same (see how the span of required disk mass between mass bin edges are linearly uniform in the right panel of Figure \ref{fig:Mpl_Mdust_relation}).\footnote{log(fx)-log(x) = log(f); log(f$^2$x) - log(fx) = log(f); and log(f$^{(n+1)}$x) - log(f$^n$x) = log(f).}
In the 2D headwind regime, $\epsilon \propto M_{\rm p}^{2/3}$ and thus $M_{\rm dust\rightarrow M_{\rm p}} \propto (M_{\rm p}^{1/3} - M_0^{1/3})$, which leads to a similar qualitative result. This dependence of $M_{\rm dust\rightarrow M_{\rm p}}$ on $M_{\rm p}$ and $M_0$ combined with the disk mass distribution sets the planet mass function at birth. For our fiducial model, we adopt a Shakura-Sunyaev viscosity parameter $\alpha_{\rm t} = 10^{-4}$ and a fragmentation velocity of dust grains $v_{\rm f} = 1$ m s$^{-1}$, but we vary these parameters to quantify their effect on the planet mass function in \S~\ref{sec:planet_mass_dist}.

Equation~\ref{eq:acc-eff} indicates that the temperature structure of the disk plays a central role in setting the accretion efficiency (through $H_{\rm g}$ and $\eta$) and the maximum pebble size (St$_{\rm frag/drift} \propto 1 / c_{\rm s}^2$). We adopt the analytical prescription for temperature structure from \cite{Ida16} based on results of \cite{Oka11} and \cite{Garaud07}:
\begin{equation}
    T_{\rm disk} = {\rm min}(2000 \, {\rm K}, {\rm max} (T_{\rm vis}, T_{\rm irr})), 
\end{equation}
where the temperature $T_{\rm vis}$ in the viscously heated region is given by (equivalent to assuming an optically thick region with a constant opacity of 0.087 cm$^2$ g$^{-1}$)
\begin{multline}
    T_{\rm vis} \simeq 200 \, {\rm K} \bigg(\frac{M_\star}{M_{\odot}}\bigg)^{3/10} \bigg(\frac{\dot{M}_\star}{10^{-8} M_{\odot} {\rm yr}^{-1}}\bigg)^{2/5} \\ \bigg(\frac{\alpha_{\rm t}}{10^{-3}}\bigg)^{-1/5} \bigg(\frac{r}{1 \, {\rm au}}\bigg)^{-9/10},
\end{multline}
and the temperature $T_{\rm irr}$ in the irradiation-dominated region is 
\begin{equation}
    T_{\rm irr} \simeq 150 {\rm K} \bigg(\frac{L_\star}{L_{\odot}}\bigg)^{2/7} \bigg(\frac{M_\star}{M_{\odot}}\bigg)^{-1/7} \bigg(\frac{r}{1 \, {\rm au}}\bigg)^{-3/7}.
\end{equation}
In this study, we are interested in stars to which microlensing surveys are most sensitive so we set $M_\star = 0.5 M_\odot$. Adopting the $\dot{M}_\star \propto M_\star^2$ dependence that is supported by observations and the median $\dot{M}_\star = 10^{-8} \, M_{\odot}$ yr$^{-1}$ for a Sun-like star \citep{Hartmann16, Manara2023}, we set $\dot{M}_\star = 2.5 \times 10^{-9} \, M_{\odot}$ yr$^{-1}$ for our $0.5 \, M_{\odot}$ star. The stellar luminosity's dependence on stellar mass varies from $d{\rm ln}L_{\star}/d{\rm ln}M_{\star} \simeq 1.4 - 1.9$ during the pre-main sequence phase and we adopt a value of 1.5 along with $L_{\star} = L_{\odot}$ for a Sun-like star \citep{Choi16, Dotter16}. With these choices and our fiducial $\alpha_{\rm t} = 10^{-4}$, the disk transitions from $T_{\rm vis}$ to $T_{\rm irr}$ at 1.5 au.

\section{Planet mass function at birth}
\label{sec:planet_mass_dist}

\begin{figure*}
    \centering
    \includegraphics[width=0.49\linewidth]{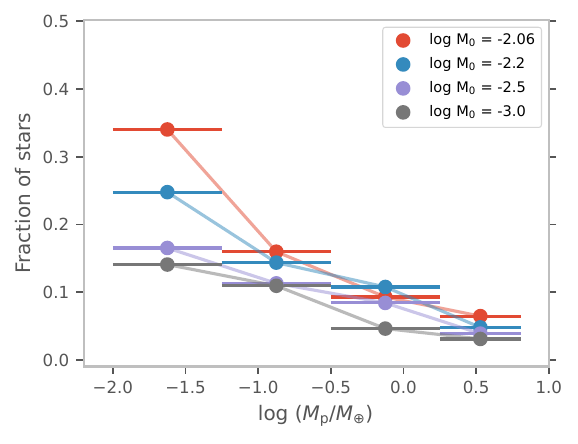}
    \includegraphics[width=0.49\linewidth]{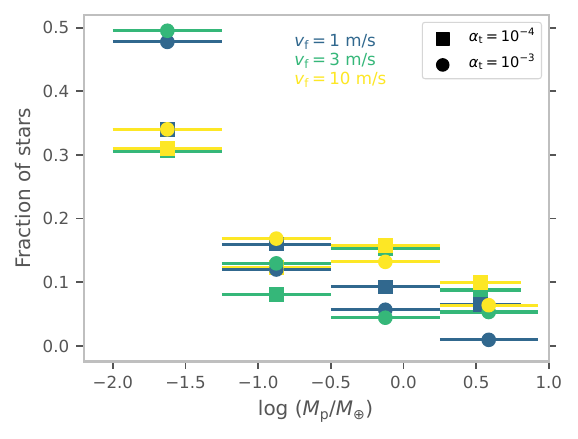}
    \caption{Fraction of stars that can grow seeds of mass $M_0$ to a given planet mass $M_{\rm p}$ at 3 au, computed by taking the difference in $f(>M_{\rm disk}=M_{\rm dust \rightarrow M_{\rm p}})$ at the minimum and maximum edge of each given $M_{\rm p}$ bin (see right panel of Figure \ref{fig:Mpl_Mdust_relation}). Left panel: This plot is for fiducial values of turbulence $\alpha_{\rm t} = 10^{-4}$ and fragmentation velocity $v_{\rm f} = 1$ m/s and the colors correspond to different initial seed masses. Right panel: Initial seed mass is fixed at log$(M_0$/M$_{\oplus}) = -2.06$ and the colors and marker indicate values of $v_{\rm f}$ and $\alpha_{\rm t}$ respectively. We expect the initial planet mass function to be more bottom-heavy for heavier seed mass, smaller $v_{\rm f}$, and larger $\alpha_{\rm t}$.}
    \label{fig:planet_mass_dist}
\end{figure*}

We calculate the fraction of stars with masses in the range of $0.4 - 0.6 \, M_{\odot}$ that have planets in a given mass bin at distances of $1-10$ au following the procedure laid out in \cite{Chachan23}. Using the pebble accretion efficiency $\bar{\epsilon}$, we estimate the dust mass $M_{\rm dust\rightarrow M_{\rm p}}$ needed to grow planets from an initial mass $M_0$ to a final mass in the range [$M_{\rm p, min}$, $M_{\rm p, max}$) where $M_{\rm p, min}$ and $M_{\rm p, max}$ delineate the limits of a given mass bin we use to construct our model planet mass function. Subsequently, we use the cumulative distribution function (CDF) of observed disk dust masses for stars in the mass range $0.4 - 0.6 \, M_{\odot}$ to obtain the fraction of stars that have enough dust mass to grow planet seeds to the given planet mass range [$M_{\rm p, min}$, $M_{\rm p, max}$). Since masses for stars that host Class I disks have not been measured, we use stellar and disk masses for Class II stars \citep{Manara2023} to build the CDF and rescale the disk masses by a factor of 3 to account for the larger disk masses observed in Class I stars. This factor is based on a comparison of the disk masses in the Class 0/I sample of \cite{Manara2023} and the Class II sample in \cite{Tobin20} (see \citealt{Chachan23} for more details) and is in agreement with previous studies that compared these two populations \citep[e.g.,][]{Tychoniec2020}.

Each of our planet mass bins span 0.75 dex in log planet mass from $10^{-2}$ M$_{\oplus}$ to the pebble isolation mass M$_{\rm iso}$ evaluated at 3 au, for a total of 4 bins. We limit ourselves to a minimum mass of $10^{-2}$ M$_{\oplus}$ as this is the smallest mass to which {\it Roman} is expected to be sensitive \citep{Penny2019, Zhu2021}. Since the minimum mass of interest is $10^{-2}$ M$_{\oplus}$, we choose a slightly lower fiducial seed mass log$(M_0$/M$_{\oplus}) = -2.06$ and vary it down to -3. This range of seed mass is motivated by the results of numerical simulations of streaming instability \citep{Simon2016, Schafer2017, Abod2019, Li2019} combined with subsequent growth of the born planetesimals \citep{Jang2022, Lorek2022}. The planetesimal mass distribution at birth is top heavy and the largest planetesimal that is produced tends to dominate subsequent pebble accretion. Our adopted range captures the range of seed masses for a 0.5 $M_\odot$ star in \cite{Liu2020} and \cite{Lorek2022}.
We broadly classify the planet mass bins as follows: Ganymede-like planets with $M_{\rm p} \in [10^{-2}, 10^{-1.25})$ M$_{\oplus}$, Mars-like planets with $M_{\rm p} \in [10^{-1.25}, 10^{-0.5})$ M$_{\oplus}$, Earth-like planets with $M_{\rm p} \in [10^{-0.5}, 10^{0.25})$, and super-Earths with $M_{\rm p} \in [10^{0.25} \, {\rm M}_{\oplus}, M_{\rm iso})$. The last bin's width can therefore be smaller than 0.75 dex if M$_{\rm iso} < 10$ M$_{\oplus}$, which is always the case for our disk parameters at 3 au. 

In Figure~\ref{fig:Mpl_Mdust_relation} (left panel), we show $M_{\rm dust\rightarrow M_{\rm p}}$ as a function of planet mass $M_{\rm p}$ for different orbital distances. We verify that $M_{\rm dust\rightarrow M_{\rm p}}$ rises linearly with log~$M_{\rm p}$ for almost the entire range of $M_{\rm p}$, as we would expect from Equation~\ref{eq:mdust_mp} in the 3D accretion regime. At higher planet masses ($M_{\rm p} \gtrsim 1$~M$_{\oplus}$) and larger orbital distances, the gradient of $M_{\rm dust\rightarrow M_{\rm p}}$ with $M_{\rm p}$ increases slightly due to a gradual shift from 3D to 2D accretion regime. In addition, the gradient of $M_{\rm dust\rightarrow M_{\rm p}}$ with $M_{\rm p}$ increases at larger orbital distances due to lower accretion efficiency further away from the star, requiring more dust mass to create a target $M_{\rm p}$.

As illustrated in the right panel of Figure~\ref{fig:Mpl_Mdust_relation}, the disk mass function is bottom-heavy for stars in the mass range $0.4 - 0.6 \, M_{\odot}$. In addition, because the required $M_{\rm dust \rightarrow M_{\rm p}}$ within a given mass bin is linearly uniform (see the discussion below equation \ref{eq:pebble_iso}), our choice of logarithmic spacing of the mass bin to construct the initial planet mass function leads to widest span in required disk mass to populate the lowest mass bin (see vertical lines in the right panel).
Visually, this span of required disk mass in the lowest $M_{\rm p}$ bin appears to narrow considerably when we choose lower $M_0$, only because the required disk mass to grow $\log M_0 = -2.5$ into $10^{-2} M_\oplus$ is much higher than starting from $\log M_0 = -2.06$ while the absolute value of difference between $M_{\rm dust \rightarrow M_{\rm p}}$ evaluated at the bin edges remains the same.

Figure~\ref{fig:planet_mass_dist} shows that the fraction of stars that produce planets in a given mass bin rises as planet mass declines, largely due to bottom-heavy disk mass function. Around a large fraction of stars, we expect planetary seeds should only grow to Ganymede and Mars mass by pebble accretion. As demonstrated in the left panel, adopting lower seed mass flattens the initial planet mass function as the fraction of stars that can nucleate Ganymede-like objects decrease substantially. Such a reduction arises because substantially more disk mass is required to grow lighter seeds to moon-mass objects, narrowing the span of $M_{\rm rightarrow M_{\rm p}}$ that correspond to the lowest $M_{\rm p}$ bin, probing less dynamic range in the disk mass CDF $f(>M_{\rm disk})$.

The effect of the turbulence strength $\alpha_{\rm t}$ and fragmentation velocity $v_{\rm f}$ on the initial planet mass function is shown in Figure~\ref{fig:planet_mass_dist}, right panel. In the 3D accretion regime, increasing $\alpha_{\rm t}$ and decreasing $v_{\rm f}$ decreases the accretion efficiency and increases the dust mass $M_{\rm dust\rightarrow M_{\rm p}}$ required to grow a planet \citep{Chachan23}, leading to the analogous effect of decreasing $M_0$ and therefore flattening the initial planet mass function. The difference in $M_{\rm dust\rightarrow M_{\rm p}}$ for two neighboring planet mass values also increases although its effect on the initial planet mass function is muted as $M_{\rm dust\rightarrow M_{\rm p}}$ is already large enough to sample the tail of disk mass CDF.

\section{How does the planet mass function evolve?}
\label{sec:planets_evolve}
\subsection{Simulation setup}
\label{sec:sim_setup}
\begin{figure}
    \centering
    \includegraphics[width=\linewidth]{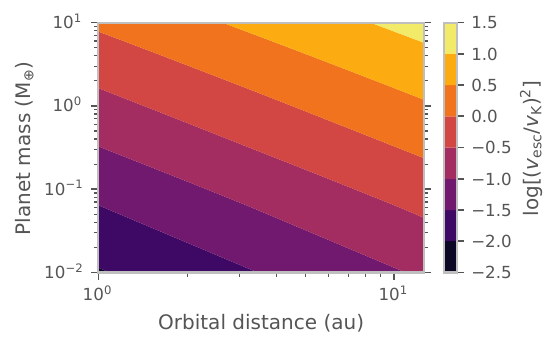}
    \caption{The ratio of a planet's surface escape velocity squared $v_{\rm esc}^2$ to its orbital velocity squared $v_{\rm K}^2$ as a function of planet's orbital distance and mass (equation \ref{eq:theta2}). For $v_{\rm esc}^2 / v_{\rm K}^2 > 1$, ejections are more likely than mergers, which we expect towards large orbital distances where the gravitational potential of the central star is shallower and for larger planet mass which can impart greater kinetic energy through scattering.}
    \label{fig:vesc_vk}
\end{figure}

In the previous section, we have quantified the fraction of stars that harbor a planet of a given mass at birth (i.e., initial coagulation). We now consider further dynamical evolution and its effect on the final observable planet mass function. Multiple protoplanets are likely to emerge in any given protoplanetary disk, and over Gyr timescale, the protoplanets are expected to undergo series of orbital instabilities that can lead to mergers, shifting population of low mass planets to higher masses. Close encounters between planets can also lead to scattering and ejection. Whether a given encounter between a pair of planets may lead to merger or ejection can be quantified with the ratio of the escape velocity at the surface of the protoplanet vs.~the circular orbital velocity of the planet \citep{Petrovich2014}:
\begin{align}
    \theta^2 &= \left(\frac{v_{\rm esc}}{v_{\rm K}}\right)^2 \nonumber \\
    &= \left(\frac{2GM_p}{R_p}\right)\left(\frac{a}{GM_\star}\right)
    \label{eq:theta2}
\end{align}
where $R_p$ is the radius of the planet set by the mass-radius relationship of \citet{Zeng2016} assuming 50:50 water-rock composition. The relation quoted by \citet{Zeng2016} does not extend below 0.0625 $M_\oplus$, so for these planets, we calculate planet radii assuming a constant density 1.89 g cm$^{-3}$ (close to the observed bulk density of Ganymede) equal to the density of the 0.0625 $M_{\oplus}$ planet. As demonstrated in Figure \ref{fig:vesc_vk}, ejection is more likely ($\theta^2 > 1$) for scattering between higher mass protocores and at larger orbital distances, as expected. Between 1 and 10 au, we would generally expect dynamical interactions between cores lighter than $\sim$0.1--1$M_\oplus$ to lead to mergers rather than ejections.

We use \texttt{REBOUND} \citep{Rein2012} to simulate a statistical ensemble of low mass planetary systems to quantify the extent to which dynamical evolution alters the planet mass function. As our starting point, we use our fiducial parameters ($\alpha_{\rm t} = 10^{-4}$, log$(M_0$/M$_{\oplus}) = -2.06$, $v_{\rm f} = 1$ m s$^{-1}$) to set the initial conditions for the N-body simulations. We envision a scenario in which planets have some initial eccentricity that is damped by disk gas for 1 Myr, our chosen disk dissipation time, after which the gas damping is removed and the planetary system is evolved to 100 Myr. At this point, we use \texttt{SPOCK} \citep{Tamayo2020b} to evaluate the stability probability of the planetary systems over $10^9$ orbits (1.4 Gyr). For some simulation sets that have a high probability of long-term instability, we extend their runtime to 1 Gyr to verify \texttt{SPOCK}'s output and to evaluate the system's long-term behavior by N-body integration to optimize our use of computational resource as 1 Gyr dynamical simulations can take a while. 

For each set of simulations, we specify four properties: the masses, separations, and initial eccentricities of the planets and the depletion factor of the disk gas that damps the planets' eccentricities for the first 1 Myr. We simulate 500 realizations for a given configuration in which the mean longitudes and arguments of pericenter are randomly drawn from a uniform distribution $U[0, 2\pi)$ and all the planets are coplanar. For the initial eccentricities of planets ($e_{\rm initial}$), the two bracketing values of $10^{-8}$ (physically representing zero; technically, nonzero value is required for numerical stability) and the disk aspect ratio $h_{\rm g} =  H_{\rm g} / r$ are chosen. 

\begin{table*}
	\begin{threeparttable}
    	\caption{Dynamical Simulations Setup} \label{table:simulations}
    	\begin{center}
        	\begin{tabular*}{\textwidth}{@{\extracolsep{\fill}} l c c c c}
        		\hline \hline
        		Simulation\textsuperscript{a} & $M_{\rm initial}$ ($M_{\oplus}$) & $e_{\rm initial}$\textsuperscript{b} & $\Delta$ ($H_{\rm g}$) & $d$ \\ \hline
        		\verb|G_3_0_1| & $10^{-1.625}$ & 0 & 3 & 1 \\
        		\verb|G_3_hg_1| & $10^{-1.625}$ & $h_{\rm g}$ & 3 & 1 \\
                \verb|G_3_hg_10| & $10^{-1.625}$ & $h_{\rm g}$ & 3 & 10\textsuperscript{c} \\
                \verb|G_3_hg_100| & $10^{-1.625}$ & $h_{\rm g}$ & 3 & 100\textsuperscript{c} \\
                \verb|GVM_3_0_1| & $10^{-2} - 10^{-1.25}$ (at 3 au) & 0 & 3 & 1 \\
                \verb|GVM_3_hg_1| & $10^{-2} - 10^{-1.25}$ (at 3 au) & $h_{\rm g}$ & 3 & 1 \\
                \verb|GVM_3_hg_10| & $10^{-2} - 10^{-1.25}$ (at 3 au) & $h_{\rm g}$ & 3 & 10 \\
                \verb|GVM_3_hg_100| & $10^{-2} - 10^{-1.25}$ (at 3 au) & $h_{\rm g}$ & 3 & 100 \\
        		\verb|G_2_0_1| & $10^{-1.625}$ & 0 & 2 & 1 \\
        		\verb|G_2_hg_1| & $10^{-1.625}$ & $h_{\rm g}$ & 2 & 1 \\
                \verb|G_2_hg_10| & $10^{-1.625}$ & $h_{\rm g}$ & 2 & 10 \\
                \verb|G_2_hg_100| & $10^{-1.625}$ & $h_{\rm g}$ & 2 & 100 \\
                \verb|M_3_0_100| & $10^{-0.875}$ & 0 & 3 & 100 \\
                \verb|M_3_hg_10| & $10^{-0.875}$ & $h_{\rm g}$ & 3 & 10 \\
                \verb|M_3_hg_100| & $10^{-0.875}$ & $h_{\rm g}$ & 3 & 100 \\
                \verb|E_3_0_100| & $10^{-0.125}$ & 0 & 3 & 100 \\
                \verb|E_3_hg_100| & $10^{-0.125}$ & $h_{\rm g}$ & 3 & 100 \\
                \verb|SE_4.5_0_100| & $10^{0.53}$ & 0 & 4.5 & 100 \\
                \verb|SE_4.5_hg_100| & $10^{0.53}$ & $h_{\rm g}$ & 4.5 & 100 \\
        		\hline
        	\end{tabular*}
        	 \begin{tablenotes}
             \small
			 \item {\bf Notes.} 
             \item \textsuperscript{a}{Simulation names indicate the planet mass bin (\verb|G|: Ganymede-like, \verb|GVM|: Ganymede-like with variable planet mass, \verb|M|: Mars-like, \verb|E|: Earth-like, \verb|SE|: super-Earth), separation in terms of scale height, initial eccentricity, and gas depletion factor.}
             \item \textsuperscript{b}{When 0, it is set to $10^{-8}$ for numerical stability. $h_{\rm g}$ is the disk aspect ratio.}
             \item \textsuperscript{c}{Simulations run to 1 Gyr.}
			\end{tablenotes}
    	\end{center}
	\end{threeparttable}
\end{table*}

For each mass bin, we take the median value to which we set all initial planet mass ($M_{\rm initial}$) in a given system to be equal (note: different from the seed mass $M_{0}$, which characterizes planet mass prior to pebble accretion). Exceptionally, for the lowest mass bin, we consider non-uniform mass in a given system. First, we sample 500 different disk masses uniformly in log $M_{\rm disk}$ from the observed CDF within range that produces planets in the lowest mass bin at 3 au (see Figure \ref{fig:Mpl_Mdust_relation}). For each of the 500 disk mass, we compute $M_{\rm initial}$ at chosen orbital distances following the location dependent pebble accretion efficiency (equation \ref{eq:acc-eff}). 

We place the innermost planet at 1 au and subsequent planets are emplaced following a chosen initial planet separation $\Delta$ until the we reach 10 au. In our fiducial set-up, we choose $\Delta =3 H_{\rm g}$. This choice is motivated by a few considerations: i) it ensures that the sum of planet masses is lower than the disk masses that are hosting such systems; ii) the typical length scale of perturbations in gas disks due to disk-planet interaction is on the order of the local disk scale height \citep{Goldreich1980, Lin1986}; iii) given that we simulate systems with initial eccentricities equal to the disk aspect ratio ($H_{\rm g} / r$), a separation of $3 \, H_{\rm g}$ likely places the systems in a configuration of marginal stability rather than dynamical isolation or assured mergers. Other values of $\Delta$ are also explored. Systems that are more tightly packed and composed of more massive planets are expected to undergo more vigorous orbital instabilities on a shorter timescale \citep[see, e.g.][and references therein]{Zhou2007, Pu2015}. For a given mass bin, if there are no mergers for a separation of 3 $H_{\rm g}$, we run simulations for a smaller separation of 2 $H_{\rm g}$. For the highest mass bin, we only run simulations for a planet separation of 4.5 $H_{\rm g}$ because we already observe vigorous planet merging and growth in the lower mass bin for a separation of 3 $H_{\rm g}$.

For our fiducial disk parameters\footnote{The other fiducial parameters, such as the fragmentation velocity of 1 m s$^{-1}$ and seed mass log$(M_0$/M$_{\oplus}) = -2.06$, set the planet mass function but do not affect the results of our N-body simulations.} ($\alpha_{\rm t} = 10^{-4}$, $\dot{M}_\star = 2.5 \times 10^{-9} \, M_{\odot}$ yr$^{-1}$) assuming steady-state accretion ($\Sigma_{\rm g} = \dot{M}_\star / 3 \pi \nu$), we find,
\begin{equation}
    \Sigma_{\rm g} \approx 5300 \, \bigg( \frac{1}{d} \bigg) \, \bigg( \frac{r}{1 \, {\rm au}} \bigg)^{-15/14} \,  {\rm g \, cm}^{-2},
\end{equation}
evaluated for the passively heated regions of the disk ($r \gtrsim 1.5$ au). The resulting $\Sigma_{\rm g}$ is comparable to the minimum-mass extrasolar nebula: $\Sigma_{\rm g} \sim 10^4$ g cm$^{-2}$ at 1 au \citep{Chiang2013}. The factor $d \in [1, 10, 100]$ quantifies the depletion factor of the disk gas, where higher $d$ implies more significant depletion.

Eccentricity damping in the first 1 Myr is computed using the `modify\_orbits' routine in \texttt{REBOUNDx} \citep{Tamayo2020}, whereby planet eccentricities are exponentially damped on an e-folding timescale of \citep{Kominami2002, Dawson16}:
\begin{align} 
\tau_{\rm d} & = \bigg(\frac{M_{\star}}{M_{\rm p}}\bigg) 
\bigg(\frac{M_{\star}}{\Sigma_{\rm g} r^2}\bigg) 
\bigg(\frac{c_{\rm s}}{v_{\rm K}}\bigg)^4 \Omega_{\rm K}^{-1} \nonumber \\ 
 & \approx  30 \, {\rm yr} \, \bigg(\frac{M_{\rm p}}{M_{\oplus}} \bigg)^{-1} \, \bigg(\frac{r}{1\,{\rm au}} \bigg)^{12/7} \,  d,
\end{align}
where the second relation has been simplified for the stellar and disk properties in the irradiated region adopted in this study ($\gtrsim$1.5 au). We use this simplified expression for all planets in our simulation domain ($1-10$ au). Our approach underestimates $\tau_{\rm d}$ for planets in the viscously heated region ($< 1.5$ au, 4 such planets for $\Delta = 3 H_{\rm g}$) by a factor of $\sim 0.55$ at most. This order unity difference is negligible compared to range of $d$ we consider and we check a posteriori that the initial effect of the error in $\tau_{\rm d}$ on the planet eccentricities is erased out by subsequent dynamical evolution. For initial eccentricities of $\sim$0, we find that varying the value of the depletion factor $d$ has no notable effect on the planet eccentricities after 1 Myr. As a result, we limit ourselves to $d=1$ for simulations with initial eccentricities of 0.

\begin{figure*}
    \includegraphics[width=\linewidth]{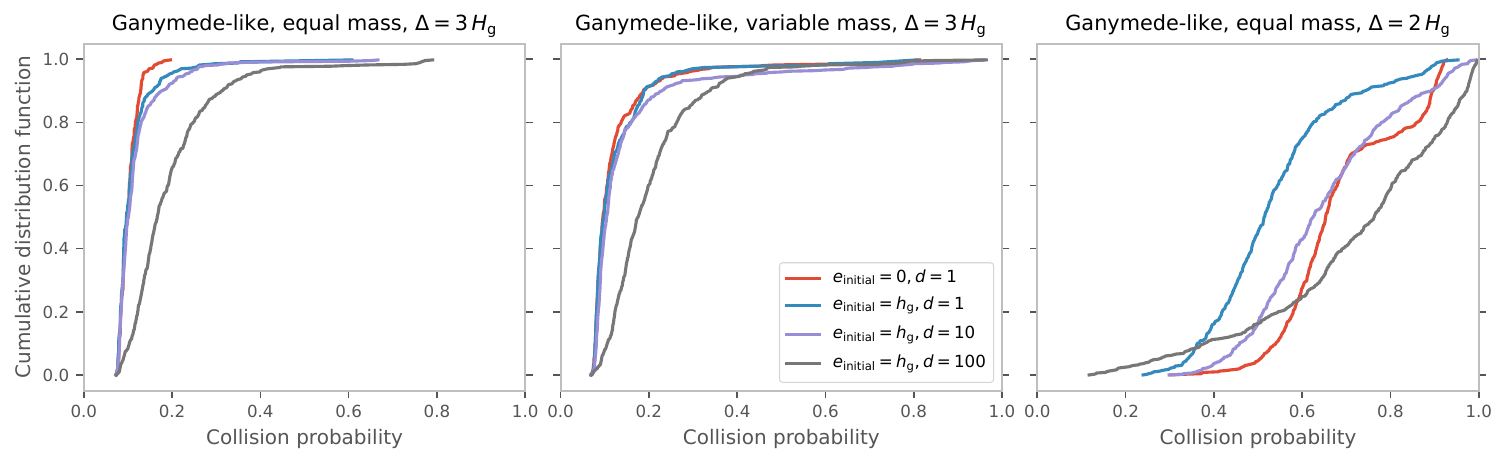}
    \caption{Cumulative distribution function of the collision probability for our simulation sets for Ganymede-like planets. We integrate our simulations for 100 Myr after which we use \texttt{SPOCK} to determine their stability probability over $10^9$ orbits. For planet separation $\Delta = 3 \, H_{\rm g}$, systems are very stable and have a low collision probability. However, for separations $\Delta \lesssim 2 \, H_{\rm g}$, the collision probability is typically higher than 0.5.}
    \label{fig:ganymede_col_prob}
\end{figure*}

Planets may migrate during the disk gas phase to the inner disk and leave the region of interest ($1-10$ au) for microlensing observations. We quantify the importance of type-I migration for different planet masses and disk gas densities by using the prescription for migration timescale $t_{\rm mig}$ from \cite{Cresswell2008}:
\begin{equation}
    t_{\rm mig} = \frac{2}{2.7 + 1.1 p} \bigg(\frac{M_\star}{M_{\rm p}}\bigg) \bigg(\frac{M_\star}{\Sigma_{\rm g} r^2}\bigg) \bigg(\frac{H_{\rm g}}{r}\bigg)^2 \Omega_{\rm K}^{-1},
\end{equation}
and integrating $\dot{a} = - a / t_{\rm mig}$ over 1 Myr, where $p = {d~\rm ln}~\Sigma_{\rm g}/{d~\rm ln}~r$.  
The migration rate is proportional to the gas surface density: a higher value of the gas depletion factor $d$ therefore implies a slower migration rate. For nearly inviscid disks (our fiducial $\alpha_{\rm t} = 10^{-4}$), feedback from piled up gas can halt migration\footnote{We note that other effects such as the unsaturated corotation torque \citep{Paardekooper2011, Kretke2012}, the presence of magnetic fields \citep{Terquem2003, Fromang2005}, and local variations in the temperature structure of the disk \citep{Benitez-Llambay2015} can also halt or reverse migration. However, we do not consider this here as such effects depend on the local complexities of the disk structure.} \citep{Rafikov2002, Fung2018} when planet mass reaches:
\begin{equation}
    M_{\rm fb} = 4 M_{\oplus} \bigg(\frac{H_{\rm g} / r}{0.035}\bigg)^3 \bigg(\frac{\Sigma_{\rm g} r^2 / M_\star}{10^{-3}}\bigg)^{5/13}.
\end{equation}
For a given planet mass bin, we limit ourselves to values of the gas depletion factor $d$ that would still retain planets in the $1-10$ au region in our setup. Planets in the lowest mass bins (`Ganymede-like') do not migrate significantly for $d \in [10, 100]$, and for $d = 1$, they stay beyond 1 au as long as their initial position $\gtrsim 2$ au. For the next mass bin (`Mars-like'), planets do not migrate significantly for $d \geq 10$. For higher mass planets (`Earth-like' and super-Earths), the combined effect of larger $t_{\rm mig}$ and lower $M_{\rm fb}$ when $d = 100$ keeps the planets from migrating significantly. Table~\ref{table:simulations} lists all the parameters and simulations that are part of this study. The properties of the simulations make up their assigned name in the following order: planet mass bin, separation in terms of disk scale height, initial eccentricity, and gas depletion factor.

\subsection{Simulation results}
\subsubsection{Ganymede-like planets}

\begin{figure}
    \centering
    \includegraphics[width=\linewidth]{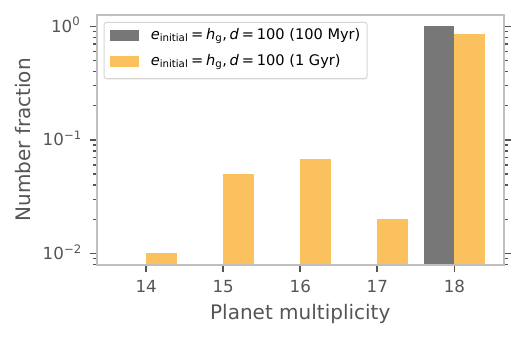}
    \caption{Planet multiplicity after 100 Myr and 1 Gyr evolution for Ganymede-like planets separated by $\Delta = 3 H_{\rm g}$ with $e_{\rm initial} = h_{\rm g}$ and $d = 100$. Only 14.8\% of systems end up having pairwise mergers when simulation runtime is extended to 1 Gyr. The final mass of the merged planets is only twice the initial value.}
    \label{fig:G_3Hg_n_pl}
\end{figure}

For our fiducial set of simulations with equal mass planets separated by 3 $H_{\rm g}$, we find that there are no mergers in all our simulations within 100 Myr. Stability analysis with \texttt{SPOCK} after 100 Myr of integration indicates the probability of collisions is very low, except for a tail of simulations with initial eccentricity $e_{\rm initial} = h_{\rm g}$ and $d = 100$ (\verb|G_3_hg_100|, Figure~\ref{fig:ganymede_col_prob}, left panel). We extend the integration time of these simulations to 1 Gyr and find that only 14.8\% of our 500 simulations have one or more pairwise mergers of the initial planets (Figure~\ref{fig:G_3Hg_n_pl}). We also integrate the simulation set \verb|G_3_hg_10| to 1 Gyr to verify \texttt{SPOCK}'s results and find that there are no mergers in this case. 

Even in case of the small fraction of planets undergoing mergers, they are all pairwise, increasing the mass of the resulting planet by a factor of two and so these planets, after merging, do not jump to the next mass bin. At a separation of 3 $H_{\rm g}$, Ganymede-mass planets therefore do not grow significantly by mergers regardless of the chosen value of $e_{\rm initial}$ and $d$.

We also test if having variable planet masses rather than equal mass planets separated by 3 $H_{\rm g}$ changes the outcome of dynamical evolution (see \S~\ref{sec:sim_setup} for details on setting the planet mass). After running the simulations for 100 Myr, we find that there are no mergers in any of the simulations. Stability analysis with \texttt{SPOCK} again indicates that collision probability in these systems over $10^9$ orbits is very low (Figure~\ref{fig:ganymede_col_prob}, middle panel). Having variable rather than equal mass planets does not change the outcome of dynamical evolution, likely because all protocores, even if their masses were varied, are small ($\lesssim 0.2$ M$_{\oplus}$). Ganymede-like planets stay as such over Gyr timescales if they are separated by $\geq 3 \, H_{\rm g}$. In addition, since these planets are small enough to undergo negligible amount of migration, we expect M dwarfs to be teeming with Ganymede-like objects at 1--10 au, at least for initial $\Delta \geq 3H_{\rm g}$.

\begin{figure}
    \includegraphics[width=\linewidth]{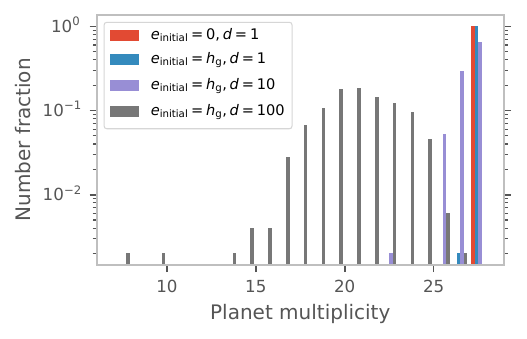}
    \includegraphics[width=\linewidth]{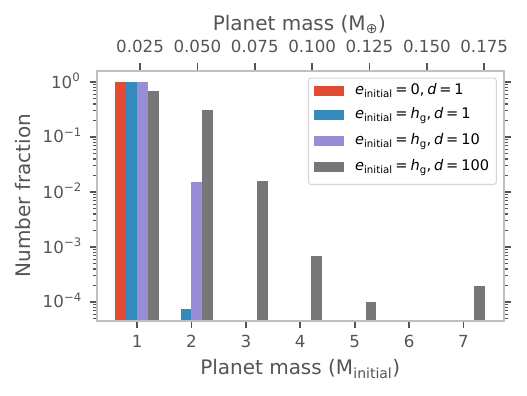}
    \caption{Planet multiplicity and mass after 100 Myr evolution for Ganymede-like planets separated by $\Delta = 2 H_{\rm g}$. When gas damping during the first 1 Myr is strong ($d = 1$), planets barely undergo any mergers after 100 Myr. For higher gas depletion and $e_{\rm initial} = h_{\rm g}$, planetary systems become dynamically unstable within 100 Myr. Nonetheless, only a small fraction of systems end up merged planets that belong to the Mars-like planetary mass bin.}
    \label{fig:G_2Hg_100Myr}
\end{figure}

Finally, to determine the fate of these planets for tighter orbital separations, we run simulations for equal mass planets spaced apart by 2 $H_{\rm g}$ (Figure~\ref{fig:G_2Hg_100Myr}). After 100 Myr of integration, we find that all of the simulations in the dynamically hottest setup (\verb|G_2_hg_100|) have had at least one merger. For $e_{\rm initial} = h_{\rm g}$ and $d = 10$, 35.2\% of our 500 simulations show at least one merger. For $d = 1$, gas damping is still effective enough to lower planet eccentricities and prevent mergers in the vast majority of our simulations (no mergers for \verb|G_2_0_1|, only one merger in \verb|G_2_hg_1|). Results from \texttt{SPOCK} indicate that most systems have a collision probability higher than 0.5 (Figure~\ref{fig:ganymede_col_prob}, right panel). The mergers in simulation set \verb|G_2_hg_10| are all pairwise and so the mass of the resulting planet is just twice the initial mass. However, for \verb|G_2_hg_100|, 169 planets (1.68\% of all planets at 100 Myr in 500 simulations) in 103 different systems (20.6\% of 500 simulations) end up with masses $\geq 3 \times M_{\rm initial}$, i.e. they end up crossing the threshold to the next mass bin and would be classified as Mars-like planets (Figure~\ref{fig:G_2Hg_100Myr}). The fraction of systems in which we expect the member planets to jump to the next mass bin by mergers over 100 Myr is still too small to significantly alter the shape of the planet mass function, especially for initially large seed mass where the initial fraction of stars harboring Ganymede-like planets is significantly higher than that of the Mars-like planets. For small seed mass $M_0$ where the initial planet mass function is more flat, the dynamical sculpting would likely be more pronounced and even cause a more peaked final mass function. Fully simulating the orbital dynamics over 1 Gyr and longer would be required to determine quantitatively the changes to the planet mass function.

\subsubsection{Mars-like planets}

\begin{figure}
    \centering
    \includegraphics[width=\linewidth]{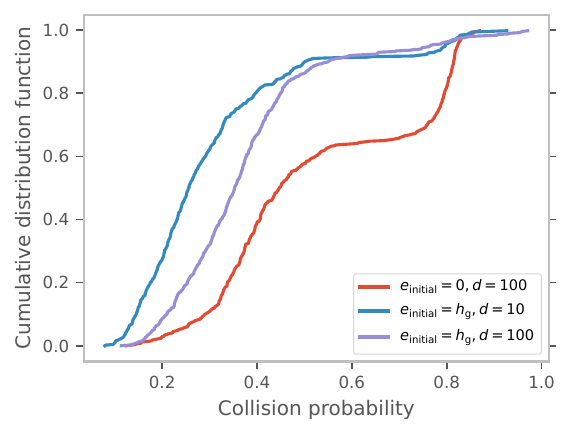}
    \caption{Cumulative distribution function of the collision probability from \texttt{SPOCK} for our simulation sets for Mars-like planets.}
    \label{fig:Mars_coll_prob}
\end{figure}

Since Ganymede-like planets already show vigorous merging when they are separated by $\Delta = 2 H_{\rm g}$ and $d = 100$, we expect Mars-like planets to undergo dynamical instabilities and merge for this separation too. We therefore set $\Delta = 3 H_{\rm g}$ and larger values of $d = 10$ and $100$ to limit ourselves to the case where migration has a negligible impact on the final location of Mars-like planets. After 100 Myr of evolution, we find that none of the systems in \verb|M_3_0_100| and \verb|M_3_hg_10| undergo any mergers. For \verb|M_3_hg_100|, one simulation has two pairwise mergers and another simulation has one pairwise merger, with no mergers seen for any of the other simulations. 

Estimates of the collision probability from \texttt{SPOCK} (Figure~\ref{fig:Mars_coll_prob}) for $e_{\rm initial} = h_{\rm g}$ indicate that 90\% of \verb|M_3_hg_10| simulations and 86\% of \verb|M_3_hg_100| simulations have a collision probability $< 0.5$. Counter-intuitively, only 58\% of \verb|M_3_0_100| simulations have a collision probability $< 0.5$ despite the dynamically colder initial conditions ($e_{\rm initial} = 0$). We find that the eccentricities of planets in the \verb|M_3_0_100| simulation suite rapidly rise after gas disk dissipation to match the eccentricities of planets in the \verb|M_3_hg_10| simulations, at times even exceeding them for the closer-in planets. Clearly, Mars-like planets dynamically stir each other up for separations $\Delta = 3 H_{\rm g}$. 

To investigate why the \verb|M_3_0_100| simulations have higher collision probabilities, we examine which `features' of the system are driving the \texttt{SPOCK} predictions. \texttt{SPOCK} calculates a number of metrics (e.g., closeness to mean-motion resonances, ratio of planet eccentricities to orbit crossing eccentricities, chaos indicators) for a given planetary system by considering nearest-planets in sets of 3 and by taking the maximum value of the metrics (i.e. the lowest stability probability) amongst all such sets. We find that the high collision probabilities for the \verb|M_3_0_100| simulations are primarily driven by the chaos indicator \texttt{MEGNO} and \texttt{MEGNOstd} \citep{Cincotta2003}, computed based on the $10^4$ yr of orbital integration. \texttt{MEGNO} measures chaos on short timescales by calculating the rate at which nearby orbits diverge. We find that for systems with $e_{\rm initial} = 0$, the rise in eccentricity is steeper with time for more of their member planets compared to simulations with $e_{\rm initial} = h_{\rm g}$ \citep[see also Figure 5 in][]{Dawson16}. This likely drives the \texttt{MEGNO} chaos indicator to higher values. 

Systems flagged as chaotic by \texttt{MEGNO} can nonetheless take a long time to go unstable depending on the time required for planet eccentricities to diffuse to orbit crossing values. To test if these systems are chaotic but not necessarily catastrophically unstable, we extend the run time of our \verb|M_3_0_100| simulations beyond 100 Myr. For 172 simulations with collision probability $< 0.5$ and 226 simulations with collision probability $> 0.5$ with runtime in the range $250-345$ Myr, no mergers are observed so at least for a few 100 Myr timescale, these systems are stable. Whether they remain so over 1 Gyr requires longer simulations. Similar to Ganymede-like objects, more than pairwise mergers are required for dynamical evolution to cause a significant change to the planet mass function between Mars-like and Earth-like mass bins.

\subsubsection{Earth-like planets}

\begin{figure}
    \centering
    \includegraphics[width=\linewidth]{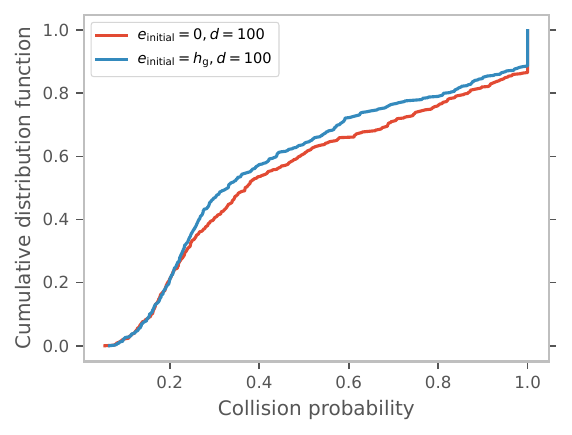}
    \caption{Cumulative distribution function of the collision probability from \texttt{SPOCK} for our simulation sets for Earth-like planets.}
    \label{fig:Earth_coll_prob}
\end{figure}

\begin{figure}
    \centering
    \includegraphics[width=\linewidth]{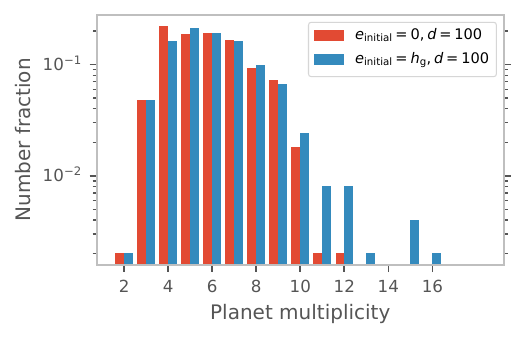}
    \includegraphics[width=\linewidth]{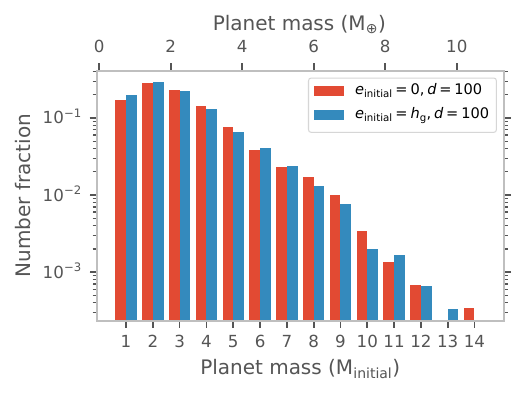}
    \includegraphics[width=\linewidth]{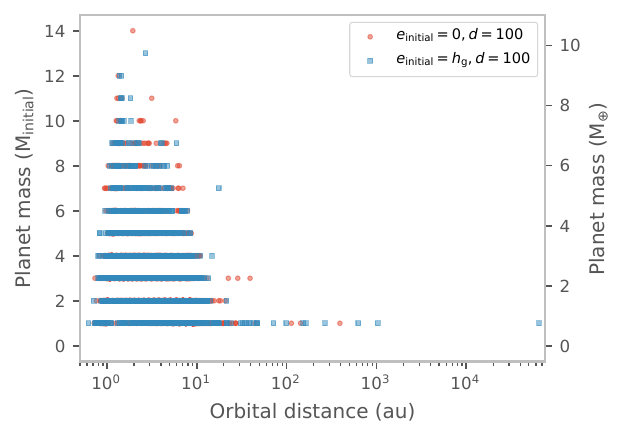}
    \caption{Histograms for planet multiplicity (top) and mass (middle) after 100 Myr evolution for Earth-like planets separated by $\Delta = 3 H_{\rm g}$. The bottom panel shows the mass and semimajor axis of all bound planets in our simulation suite at $t = 100$ Myr.}
    \label{fig:E_3Hg_100Myr}
\end{figure}

Both Ganymede and Mars-like planets do not undergo large scale mergers for separations $\Delta \geq 3 H_{\rm g}$. We therefore choose a value of $\Delta = 3 H_{\rm g}$ for Earth-like planets to determine if dynamical evolution at such planet separations significantly affects their population. Figures~\ref{fig:Earth_coll_prob} and \ref{fig:E_3Hg_100Myr} show the collision probability calculated from \texttt{SPOCK} and the planet multiplicity, mass and semimajor axis distribution after 100 Myr. We exclude simulations (5 from \verb|E_3_hg_100|, 1 from \verb|E_3_0_100|) in which a planet gets scattered to semi-major axis $< 0.1$ au as we do not evolve close-in planets that would prohibitively extend the simulation timescale. 

Figures~\ref{fig:Earth_coll_prob} and \ref{fig:E_3Hg_100Myr} show that the state of these systems after 100 Myr is weakly dependent on the initial eccentricities of the planets, suggesting high degree of orbital instabilities and series of mergers in both cases. The $e_{\rm initial} = h_{\rm g}$ simulations have slightly more systems at higher multiplicity, implying that these systems are marginally dynamically colder, in agreement with the collision probabilities CDF as well. We see that the final planet multiplicity peaks at $4-7$ planets per system compared to the initial multiplicity of 18 with the most common planet mass being twice the initial mass. For \verb|E_3_0_100| simulations, 1598 of the 2935 planets (54.4\%) remaining at 100 Myr in 498 out of 499 systems have masses $\geq 3 M_{\rm initial}$, i.e. they would be qualified as super-Earths, jumping to the next mass bin. Similarly, for \verb|E_3_hg_100| simulations, we find that 1545 of the 3031 planets (50.1\%) remaining at 100 Myr in 490 out of 495 systems have masses $\geq 3 M_{\rm initial}$. Further evolution will likely increase the fraction of Earth-like planets that would jump to the next mass bin (39.2\% of \verb|E_3_0_100| and 36.0\% \verb|E_3_hg_100| simulations have collision probability $> 0.5$). Earth-like planets separated by $\Delta = 3 H_{\rm g}$ are therefore likely to be heavily dynamically sculpted, potentially leading to a gap in the final planet mass function at $\sim$Earth mass. The mass at which such a gap appears, if observed, can inform the initial orbital separations. Observation of a significant population of Earth-like planets would imply wider initial separations. The gap in the mass function at Earth-like planet masses could partially be filled in by mergers of Mars-like planets if they are born separated by $\Delta \lesssim 2 H_{\rm g}$.

The bottom panel of Figure~\ref{fig:E_3Hg_100Myr} shows the mass and orbital semimajor axis of all bound planets at 100 Myr (note: it does not show which planets are in the same system). Low mass planets are scattered out far beyond 10 au on highly eccentric orbits, at times to such large distances that the planets would \textit{appear} to be free-floating although physically bound. We also find truly unbound planets with eccentricities $> 1$. In the \verb|E_3_0_100| simulations, 149 out of 2935 (5.1\%) total planets at 100 Myr are unbound, of which 86 retain their initial mass. The corresponding number for \verb|E_3_hg_100| simulations is 121 out of 3031 (4\%) planets, with 64 planets that retain their initial mass. The observed dynamical interactions of Earth-like planets that lead to ejections and large scale scattering is in line with expectations from Figure~\ref{fig:vesc_vk}.

\subsubsection{Super-Earths}

\begin{figure}
    \centering
    \includegraphics[width=\linewidth]{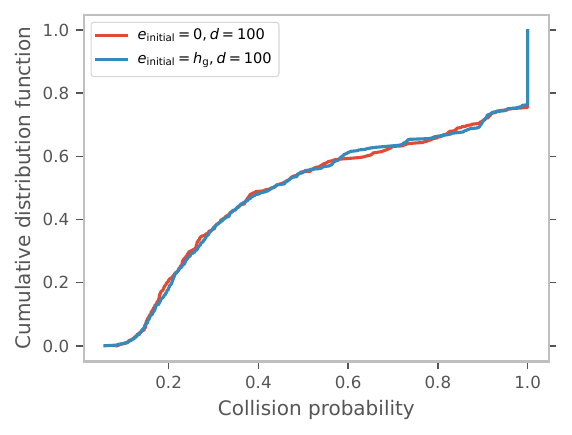}
    \caption{Cumulative distribution function of the collision probability from \texttt{SPOCK} for our simulation sets for super-Earths.}
    \label{fig:SE_coll_prob}
\end{figure}

\begin{figure}
    \centering
    \includegraphics[width=\linewidth]{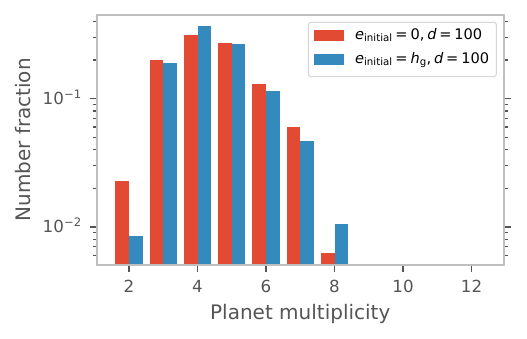}
    \includegraphics[width=\linewidth]{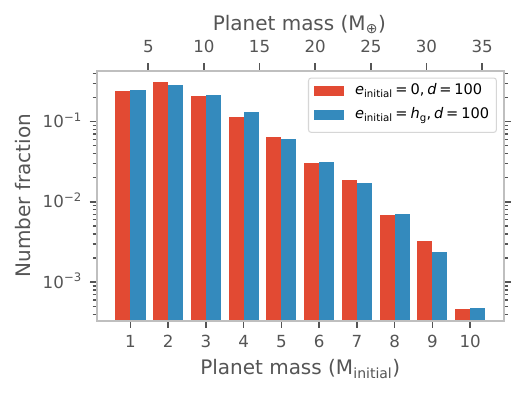}
    \includegraphics[width=\linewidth]{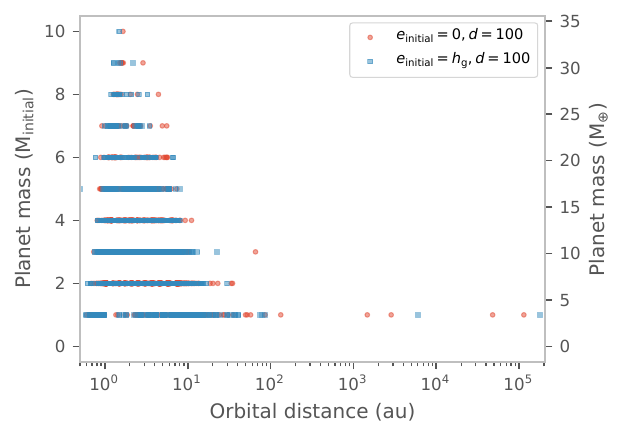}
    \caption{Histograms for planet multiplicity (top) and mass (middle) after 100 Myr evolution for super-Earths separated by $\Delta = 4.5 H_{\rm g}$. The bottom panel shows the mass and semimajor axis of all bound planets in our simulation suite at $t = 100$ Myr.}
    \label{fig:SE_4.5Hg_100Myr}
\end{figure}

Since Earth-like planets merge vigorously for separation $\Delta = 3 H_{\rm g}$, we simulate super-Earths with a larger separation of $\Delta = 4.5 H_{\rm g}$. This separation lies in between the bracketing values obtained by scaling $\Delta$ for super-Earths from Earth-like planets by using the relations for dynamical separation from \cite{Hadden2018} ($\Delta \propto M_{\rm p}^{1/4}$) and separation in terms of Hill radius ($\Delta \propto M_{\rm p}^{1/3}$, \citealt{Gladman1993, Petit2018}). In the following analysis, we again exclude simulations (26 from \verb|SE_3_hg_100|, 17 from \verb|SE_3_0_100|) in which a planet gets scattered to semi-major axis $< 0.1$ au. Figures~\ref{fig:SE_coll_prob} and \ref{fig:SE_4.5Hg_100Myr} show the collision probability calculated from \texttt{SPOCK} and the planet multiplicity, mass, and semimajor axis distribution at 100 Myr. We see that the memory of initial eccentricity is even more erased as compared to Earth-like planet case suggesting vigorous dynamical sculpting by planet-planet interactions.

This vigorous sculpting is evidenced by the sharper peak of planet multiplicity at $\sim$4 (top panel of Figure \ref{fig:SE_4.5Hg_100Myr}) compared to Earth-like planets at $\Delta = 3 H_{\rm g}$. Like Earth-like planet case, most mergers are pairwise leading to the modal planet mass being twice the initial mass after 100 Myr. For \verb|SE_3_0_100| simulations, we find that 972 of the 2168 planets (44.8\%) remaining at 100 Myr in 471 out of 483 systems have masses $\geq 3 M_{\rm initial}$, large enough to jump to the next mass bin. Similarly, for \verb|SE_3_hg_100| simulations, 980 of the 2120 planets (46.2\%) remaining at 100 Myr in 467 out of 474 systems have masses $\geq 3 M_{\rm initial}$. Mergers of super-Earths can therefore easily produce planets that are akin to Neptune. Further evolution will likely increase the fraction of super-Earths that grow beyond $M_{\rm iso}$ and evacuate the initial mass bin (44.9\% of systems for sets of simulations have collision probability $> 0.5$). Combined with most systems with Earth-like planets undergoing mergers, we would expect overall shift of the initial mass function for Earth mass and up towards higher masses, unless the initial $\Delta > 4.5 H_{\rm g}$.

The mass and orbital semimajor axis of all bound planets at 100 Myr are shown in the bottom panel of Figure~\ref{fig:SE_4.5Hg_100Myr} (note: it does not show which planets are in the same system). Similar to the case for Earth-like planets, super-Earths that retain their initial mass are scattered out far beyond 10 au on highly eccentric orbits and would ostensibly be considered free-floating. The fraction of unbound planets is higher for super-Earths, as one might expect from Figure~\ref{fig:vesc_vk}. For the \verb|SE_3_0_100| simulations, 244 out of 2168 (11.3\%) total planets at 100 Myr are unbound, of which 165 have a mass $= M_{\rm initial}$. For the \verb|SE_3_hg_100| simulations, 224 out of 2120 (10.6\%) planets are unbound, of which 157 planets that have a mass of $M_{\rm initial}$.

\section{Discussion and Conclusion}
\label{sec:discussion}

The \emph{Nancy Grace Roman Space Telescope} will enable us to find Ganymede-mass to Earth-mass objects at $1-10$ au for the first time. The sensitivity of the telescope to the lowest planet masses is however low and therefore the number of Ganymede and Mars-like objects that we expect to find is highly sensitive to the underlying planet mass function.

Combining mass growth by pebble accretion with the observed disk mass function, we find that:
\begin{enumerate}
    \item The initial planet mass function is generally bottom-heavy and increasingly so for larger initial seed mass approaching that of Ganymede.
    \item Consequent orbital evolution for a few 100 Myr expect minimal change to the mass function for $M_{\rm p} \leq 10^{-0.5} M_\oplus$ and vigorous instabilities and mergers that lead to a potential gap at $\sim$1$M_\oplus$, if the planets were initially spaced at $\Delta = 3 H_{\rm g}$. The location of this gap is expected to shift to lower mass for a tighter initial separation and to higher mass for a wide initial separation.
    \item For planets with $M_{\rm p} \gtrsim M_\oplus$, we expect a small fraction of planets ($\sim$4--5\% Earth-like and $\sim$11\% for super-Earths among the total simulated planets) being scattered to wide orbits ($\gtrsim$100 au) or become unbound (with the unbound population dominating over wide-orbit population by two orders of magnitude at least at 100 Myr), contributing to the population of free-floating planets.
\end{enumerate}

Our findings suggest that the shape of the observed planet mass function can be used to infer the initial orbital architecture of the planetary systems. More steeply bottom-heavy mass function would imply higher initial seed mass and if we find a signature of a gap in the mass function, the location of the gap can be leveraged to infer the initial planet-planet spacing.

Ejection from bound orbits by dynamical scattering is easier for systems with higher mass objects (see Figure \ref{fig:vesc_vk}). Given our setup of the problem where all planets within a given system start with the same mass, we would expect the resulting mass function of free-floating planets that is either top-heavy or flat beyond $\gtrsim$1$M_\oplus$ (with a sharp decline below), with the exact mass being dependent on the initial planet separations. Measurements of the relative abundance of bound and free-floating planets and the planet mass function of free-floating planets by \emph{Roman} will therefore serve as a probe of the primordial dynamical separation of massive ($\gtrsim$ 1 M$_\oplus$) planets.

Our work generally expects a bottom-heavy bound planet mass function and top-heavy free-floating planet mass function, which is potentially at odds with the current ground-based microlensing surveys \citep[see][their Figure 6]{Sumi2023}. Under our calculations, starting with a significantly lighter initial seed mass followed by dynamical evolution can reduce the population of Ganymede-like objects compared to that of Mars-like objects, driving the bound planet mass function to be more top-heavy. Tighter initial orbital spacing ($\Delta <3 H_{\rm g}$) could further evacuate the population of Mars-like objects, adding to the population of their more massive counterparts. Creating a bottom-heavy free-floating planet mass function may be possible if the systems with Ganymede and Mars-like objects also harbor more massive planets ($\gtrsim$1$M_\oplus$) and/or orbited by a binary companion that can dynamically eject these small objects \citep{Coleman2024}. Given that the observed disk mass fraction is bottom-heavy, most disks that can nucleate objects of $\sim$Mars mass and below do not have enough material to also create Earth-like and more massive planets. It is possible that such low mass disks do not even create the initial seed mass such that they should not be considered as planet-forming disks \citep[see e.g.][for a condition on local dust-to-gas ratio to trigger clumping]{Li2021}. In this case, the large population of free-floating low mass objects would have originated from more massive disks as part of a family with more massive planets. Whether the free-floating planet mass function is truly bottom-heavy or not remains to be verified with space-based missions.


\section*{Acknowledgments}
This research was enabled in part by support provided by (Calcul Qu\'ebec) and the Digital Research Alliance of Canada (alliance.can..ca). Y.C. acknowledges support from the Natural Sciences and Engineering Research Council of Canada (NSERC) through the CITA National Fellowship and the Trottier Space Institute through the TSI Fellowship. Y.C. is grateful for discussions with Eugene Chiang, Sam Hadden, and Yanqin Wu at the 2023 CITA Planet Day Meeting that inspired this work. E.J.L. gratefully acknowledges support by NSERC, by FRQNT, by the Trottier Space Institute, and by the William Dawson Scholarship from McGill University.

%

\vspace{5mm}


\software{astropy \citep{2013A&A...558A..33A, 2018AJ....156..123A},
          \texttt{REBOUND} \citep{Rein2012},
          \texttt{REBOUNDx} \citep{Tamayo2020},
          \texttt{SPOCK} \citep{Tamayo2020b},
          Matplotlib \citep{Hunter2007},
          Numpy \citep{harris2020array}
          }





\bibliography{sample631}{}
\bibliographystyle{aasjournal}


\end{document}